# Assuring the emotional and cultural intelligence of intelligent software systems

Alvine, ABB, Boaye Belle

Lassonde School of Engineering, York University, Toronto, Canada, alvine.belle@lassonde.yorku.ca

Intelligent software systems (e.g., conversational agents, profiling systems, recruitment systems) are often designed in a manner which may perpetuates anti-Black racism and other forms of socio-cultural discrimination. This may reinforce social inequities by supporting the automation of consequential and sometimes unfair decisions that may be made by such systems and which may have an adverse impact on credit scores, insurance payouts, and even health evaluations, just to name a few. My lightning talk will therefore emphasize the need to propose a new type of non-functional requirements called ECI (emotional and cultural intelligence) requirements that will aim at developing discrimination-aware intelligent software systems. Such systems will notably be able to behave empathetically toward everyone, including minoritized groups and will ensure they are treated fairly. My talk will also emphasize the need to develop novel system assurance solutions to assure these ECI requirements are sufficiently supported by intelligent software systems.

CCS CONCEPTS • Software and its engineering -> Software organization and properties -> Extra-functional properties

## 1 INTRODUCTION

When developing intelligent systems, computing professionals (e.g., software developers, software testers) may not always be well-equipped to make them discrimination-aware by notably making them able to better capture and fairly process socio-cultural expressions, to better handle cultural differences, to avoid automating biased and unfair human decisions. This may make such systems behave in a way that lacks empathy, reinforces stereotypical behaviors, promote prejudices, and the like [1]. This could lead to the creation of discriminating intelligent systems that are not able to properly tackle digital discrimination [1]. This may reinforce social inequities by supporting the implementation of consequential and sometimes unfair decisions that may be made by such systems and which may have an adverse impact on credit scores, insurance payouts, health evaluations, etc. [2]. It is therefore crucial to contribute to the development of solutions aiming at creating discrimination-aware intelligent software systems that will behave more empathetically toward everyone, including people from underrepresented groups (e.g., Black people). One of such solutions may include: 1) the proposal of a new type of non-functional requirements called *ECI (Emotional and Cultural Intelligence)* requirements that should be supported by intelligent software systems; and 2) the development of system assurance techniques allowing to determine if these non-functional requirements are sufficiently supported by the developed systems.

## 2 BACKGROUND CONCEPTS: ASSURANCE OF SYSTEM REQUIREMENTS

An assurance case can be defined as: "*a reasonable, auditable argument created to support the contention that a defined system will satisfy particular requirements, along with supporting evidence*" [3]. Assurance cases are used in several application domains (e.g., automotive, aerospace, healthcare) to prove to stakeholders (e.g., regulatory bodies) that certain non-obvious properties (e.g., security, safety) are present in the system at hand. The goal structuring notation (GSN) is the most-used graphical notation to represent assurance cases. GSN diagrams are aligned with the concepts of the SACM (Structured Assurance Case Metamodel) that OMG issued to promote standardization and interoperability.

## 3 ASSURANCE OF EMOTIONAL AND CULTURAL INTELLIGENCE REQUIREMENTS

Hofstede [4] defined culture as "*The collective programming of the mind which distinguishes the members of one human group from another. . .*". Thus, culture is the glue that bounds together people belonging to the same social group through a set of well-agreed beliefs, norms and values. Culture has both conscious and unconscious dimensions. Culture is therefore a shared mental hallucination, an abstraction of the mind that finds its grounds in day-to-day interactions and that is systemically passed along generations within the same social group. Culture preconditions people since it surrounds them since their inception. Culture discriminates each social group from the other: it is distilled in the way people of each social group pray, study, eat, talk, etc. [5]. Hence, the same information, can be interpreted differently by people from different cultures, resulting in different positive or negative associations. Such perception barriers might lead to misunderstandings, to conflicts (e.g., racial conflicts or social injustices), which hinders the understanding and acceptance of the other who is nothing more and nothing less than our alter ego.

Still, culture is a dynamic, adaptive notion that is doomed to evolve when exposed to other social groups with different cultural backgrounds provided they are willing to bridge negative associations induced by perception barriers. That evolution is therefore possible if people from different social groups are willing to address potential cultural-related conflicts by leveraging their respective emotional and cultural intelligences to learn more about others, to eventually realize that, at the end of the day, they share common purposes. This is the reason why Paulo Coehlo [6] stated that: "*Culture makes people understand each other better. And if they understand each other better in their soul, it is easier to overcome […] barriers. But first they have to understand that their neighbour is, in the end, just like them, with the same problems, the same questions*". But how to represent, assure, or even "teach" emotional and cultural intelligence (ECI) to intelligent software systems throughout their lifecycle and prevent such systems to automate biased/unfair human decisions? By notably relying on assurance cases, of course! Using assurance cases to assure the presence of ECI capabilities in intelligent software systems could help mitigate anti-Black racism and other forms of social inequities by being a means to assure ECI requirements in intelligent systems throughout their lifecycle. My lightning talk will provide some guidance on how to leverage existing cultural frameworks (e.g., Hofstede's cultural framework) to derive emotional and cultural intelligence requirements that embody concepts such as self-awareness, self-management, social awareness and relationship management among others. My talk will also provide more guidance on how to develop system assurance (e.g., assessment, design, refactoring) techniques to yield robust discrimination-aware intelligent software systems.